\documentclass[preprint,amsmath,amssymb,superscriptaddress,notitlepage]{revtex4-1}
\usepackage{epsfig}
\usepackage{graphicx}
\usepackage{color}
\usepackage{bm}
\usepackage{siunitx}

\begin{document}

\title{Imaging and writing magnetic domains in the non-collinear antiferromagnet Mn$_{\text{3}}$Sn}

\author{Helena~Reichlova}
\affiliation{Institut f{\"u}r Festk{\"o}rper- und Materialphysik and W{\"u}rzburg-Dresden Cluster of Excellence ct.qmat, Technische Universit{\"a}t Dresden, 01062 Dresden, Germany}
\author{Tomas~Janda}
\affiliation{Faculty of Mathematics and Physics, Charles University, Ke Karlovu 3, 121 16 Prague 2, Czech Republic}
\author{Joao~Godinho}
\affiliation{Institute of Physics ASCR, v.v.i., Cukrovarnick\'a 10, 162 53, Praha 6, Czech Republic}
\affiliation{Faculty of Mathematics and Physics, Charles University, Ke Karlovu 3, 121 16 Prague 2, Czech Republic}
\author{Anastasios~Markou}
\affiliation{Max Planck Institute for Chemical Physics of Solids, N\"othnitzer Stra\ss e 40, 01187 Dresden, Germany}
\author{Dominik~Kriegner}
\affiliation{Max Planck Institute for Chemical Physics of Solids, N\"othnitzer Stra\ss e 40, 01187 Dresden, Germany}
\affiliation{Institute of Physics ASCR, v.v.i., Cukrovarnick\'a 10, 162 53, Praha 6, Czech Republic}
\author{Richard~Schlitz}
\affiliation{Institut f{\"u}r Festk{\"o}rper- und Materialphysik and W{\"u}rzburg-Dresden Cluster of Excellence ct.qmat, Technische Universit{\"a}t Dresden, 01062 Dresden, Germany}
\author{Jakub~Zelezny}
\affiliation{Institute of Physics ASCR, v.v.i., Cukrovarnick\'a 10, 162 53, Praha 6, Czech Republic}
\author{Zbynek~Soban}
\affiliation{Institute of Physics ASCR, v.v.i., Cukrovarnick\'a 10, 162 53, Praha 6, Czech Republic}
\author{Mauricio~Bejarano}
\affiliation{Helmholtz-Zentrum Dresden-Rossendorf, Institute of Ion Beam Physics and Materials Research,
Bautzner Landstra\ss e 400, 01328 Dresden, Germany}
\author{Helmut~Schultheiss}
\affiliation{Helmholtz-Zentrum Dresden-Rossendorf, Institute of Ion Beam Physics and Materials Research,
Bautzner Landstra\ss e 400, 01328 Dresden, Germany}
\author{Petr~Nemec}
\affiliation{Faculty of Mathematics and Physics, Charles University, Ke Karlovu 3, 121 16 Prague 2, Czech Republic}
\author{Tomas~Jungwirth}
\affiliation{Institute of Physics ASCR, v.v.i., Cukrovarnick\'a 10, 162 53, Praha 6, Czech Republic}
\affiliation{School of Physics and Astronomy, University Of Nottingham, NG7 2RD, Nottingham, United Kingdom}
\author{Claudia~Felser}
\affiliation{Max Planck Institute for Chemical Physics of Solids, N\"othnitzer Stra\ss e 40, 01187 Dresden, Germany}
\author{Joerg~Wunderlich}
\affiliation{Institute of Physics ASCR, v.v.i., Cukrovarnick\'a 10, 162 53, Praha 6, Czech Republic}
\affiliation{Hitachi Cambridge Laboratory, Cambridge CB3 0HE, United Kingdom}
\author{Sebastian~T.~B.~Goennenwein}
\affiliation{Institut f{\"u}r Festk{\"o}rper- und Materialphysik and W{\"u}rzburg-Dresden Cluster of Excellence ct.qmat, Technische Universit{\"a}t Dresden, 01062 Dresden, Germany}



\maketitle
\pagebreak
Harnessing the unique properties of non-collinear antiferromagnets (AFMs) will be essential for exploiting the full potential of antiferromagnetic spintronics \cite{Jungwirth2018,Baltz2018}. Indeed, many of the effects enabling ferromagnetic spintronic devices have a corresponding counterpart in materials with non-collinear spin structure \cite{Suergers2014,Zhang2014,Nakatsuji2015,Nayak2016,Ikhlas2017,Higo2018}. In addition, new phenomena such as the magnetic spin Hall effect \cite{Kimata2019} or the chiral anomaly \cite{Kuroda2017} were experimentally observed in non-collinear AFMs, and the presence of the equivalent to the ferromagnetic spin transfer torque via spin polarized currents was theoretically predicted \cite{Zelezny2017}.  In spite of these developments, an interpretation of the rich physical phenomena observed in non-collinear antiferromagnets is challenging, since the microscopic spin arrangement, the magnetic domain distribution, and the domain orientations have proven notoriously difficult to access experimentally. 

This is all the more problematic, as imaging and writing magnetic domains is of central importance for applications. Successful imaging is a basic requirement to experimentally confirm the spin transfer torque acting on non-collinear domain walls and therefore of eminent interest. 
Here, we demonstrate that the local magnetic structure of the non-collinear AFM Mn$_{\text{3}}$Sn films can be imaged by scanning thermal gradient microscopy (STGM) \cite{Weiler2012}. The technique is based on scanning a laser spot over the sample's surface, and recording the ensuing thermo-voltage. We image the magnetic structure at a series of different temperatures and show that at room temperature, the domain structure is not affected by the application of moderate magnetic fields. In addition to imaging, we establish a scheme for heat-assisted magnetic recording, using local laser heating in combination with magnetic fields to intentionally write domain patterns into the antiferromagnet. 

The antiferromagnetic semimetal Mn$_{\text{3}}$Sn is a prime representative of materials with a triangular spin structure and it is very actively discussed in the context of Weyl physics\cite{Kuebler2014,Yang2017,Ikhlas2017,Kuroda2017,Liu2017,Manna2018}. The material is of particular interest due to the topology of its electronic bands with strong Berry curvature contributions to anomalous magneto-transport \cite{Nayak2016}. It has hexagonal structure with $P$6$_3/mmc$ space group with the magnetic moments residing in a $c$-plane kagome lattice (Fig. 1a). Figure 1 shows the spin configuration experimentally confirmed in bulk Mn$_{\text{3}}$Sn \cite{Tomiyoshi1982}, but several spin arrangements in the $c$-plane are discussed as energetically equivalent \cite{Sticht1989,Brown1990,Zhang2013a,Sung2018}. Considering that our films exhibit the same N\'eel temperature $T_{\text{N}}$=420~K as bulk Mn$_{\text{3}}$Sn, we expect that they also have an identical spin structure. The anisotropy of the three magnetic sublattices partly cancels out \cite{Tomiyoshi1982} which allows for the experimental manipulation of the antiferromagnetic order.

We now discuss the origin of the signal in our thermal gradient microscopy. From a symmetry point of view, the anomalous Hall effect (AHE) is equivalent to a time-reversal odd axial vector \textbf{g} \cite{Kleiner1966,Smejkal2019} such that anomalous Hall current $j_\text{AHE}$~=~\textbf{g}$\times$\textbf{E} where \textbf{E} is the electric field. The anomalous magneto transport plane (the plane in which the electric fields or currents are applied and recorded) is then perpendicular to \textbf{g}. Considering the Mott relation the symmetry of the anomalous Nernst effect (ANE) and AHE is identical \cite{Guo2017} with the applied electric field \textbf{E} replaced by a thermal gradient $\nabla T$: $V_\text{ANE}\sim$ $\nabla T \times$\textbf{g}. The orientation of the vector \textbf{g} is determined by the non-collinear structure, as illustrated in Fig. 1b for two opposite domains. Note that the symmetry properties of the vector \textbf{g} also result in a tendency of the material to develop a net magnetic moment along the \textbf{g} vector. This net moment is, however, very weak ($\sim$ 0.002 $\mu_{\text{B}}$/f.u \cite{Nakatsuji2015}) and is not the source of the strong anomalous transport coefficients \cite{Nakatsuji2015,Smejkal2019} in Mn$_{\text{3}}$Sn.

The Mn$_{\text{3}}$Sn epitaxial thin films studied here were prepared by UHV sputtering \cite{Markou2018} (details on fabrication and characterization are compiled in the Methods and SI). The films
are oriented, such that the [001] direction ($c$-axis) points out-of-plane and 5~$\mu$m wide Hall bars patterned by optical lithography are typically oriented along the [100] direction of Mn$_{\text{3}}$Sn. A typical device and the experimental geometry are shown in Fig. 1c,d. The sample is placed in an optical cryostat with the magnetic field applied along the $x$ direction (perpendicular to the Hall bar).

We scan a focused red laser ($\lambda$=800~nm, $P$=10~mW) across the Hall bar and record the ensuing thermo-voltage along the $y$ direction (Fig.~1c). As evident from Fig.~1e, the spatially resolved thermo-voltage response clearly reveals spatial contrast.

\begin{figure}[h]
\hspace*{0cm}\epsfig{width=1\columnwidth,angle=0,file=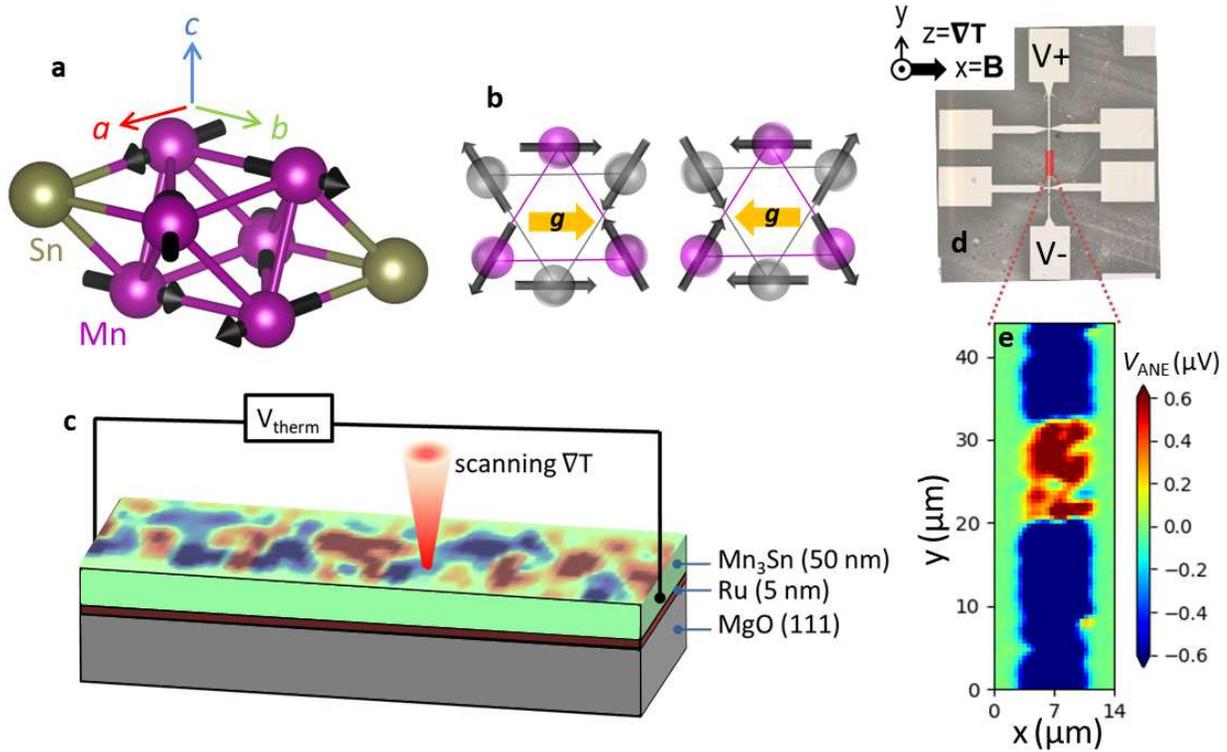}
\caption{Crystal structure and experimental setup. \textbf{a} Mn$_{\text{3}}$Sn  has hexagonal structure with $P$6$_3/mmc$ space group. The magnetic moments (spins) are residing in the $c$-plane. \textbf{b} Two opposite magnetic domains with the corresponding \textbf{g} vectors (yellow arrows). \textbf{c} Schematics of STGM. Mn$_{\text{3}}$Sn films are grown on MgO (111) substrates with a 5 nm Ru underlayer. The laser beam is scanned over the sample surface and the resulting local thermo-voltage sign and magnitude reflects the local magnetic properties. \textbf{d} Microscope image of a typical device and the experimental geometry. \textbf{e} Image of the spatially resolved thermo-voltage recorded in the sample (in the part highlighted as a red rectangle in (\textbf{d})). The thermo-voltage signal reveals clear magnetic contrast. The experiment was performed at 300~K, with a laser power of 10~mW, and in zero external magnetic field.}
\label{f1}
\end{figure}

As demonstrated by Ikhlas et al. \cite{Ikhlas2017}, the anomalous Nernst response of Mn$_{\text{3}}$Sn is strongly anisotropic. No anomalous Nernst voltage $V_\text{ANE}$ is expected when the thermal gradient $\nabla T$ is applied in the $c$-plane \cite{Ikhlas2017}, while $V_\text{ANE}$ in the direction perpendicular to the $\textbf{g}$ vector  is expected when $\nabla T$ is applied along the $c$-axis. In our sample, laser generated heat drains into the substrate, the in-plane components (in $c$-plane) compensate each other and the remaining thermal gradient is along the $z$ direction ($c$-axis). Magnetic field is applied along the $x$ direction and we detect $V_\text{ANE}$ along the $y$ direction. Thus, the thermo-voltage signal is determined by the projection of the $\textbf{g}$ vector onto the $x$ direction. The different magnitude and sign of $V_\text{ANE}$ observed depending on the position $(x,y)$ of the laser spot on the sample thus reflects the corresponding local orientation of the vector $\textbf{g}$ in the irradiated area. Note that the laser beam is focused to a diameter of 1.5~$\mathbf{\mu}$m, such that $V_\text{ANE}$ reflects the average of possibly different $\textbf{g}$ orientations or domains present within the illuminated spot. 
The measured voltage is therefore proportional to the net (average) component of the \textbf{g} vectors perpendicular to the detection direction. The resolution of the STGM is discussed in more detail in the SI. Remarkably, we observe very different $V_\text{ANE}$ patterns depending on the sample history. As discussed below, magnetic domains can be intentionally written into the Mn$_{\text{3}}$Sn using local heating. The domain pattern shown in Fig. 1e is the result of such a  process. On the other hand, upon cooling the sample from $T>T_\text{N}$ in zero magnetic field, domains are randomly populated resulting in more complex magnetic spatial contrast, as can be seen for example in Fig.~2a,b. Several Hall bars with different orientation with respect to the [100] direction of Mn$_{\text{3}}$Sn were measured showing similar results.

\begin{figure}[h]
\hspace*{0cm}\epsfig{width=1\columnwidth,angle=0,file=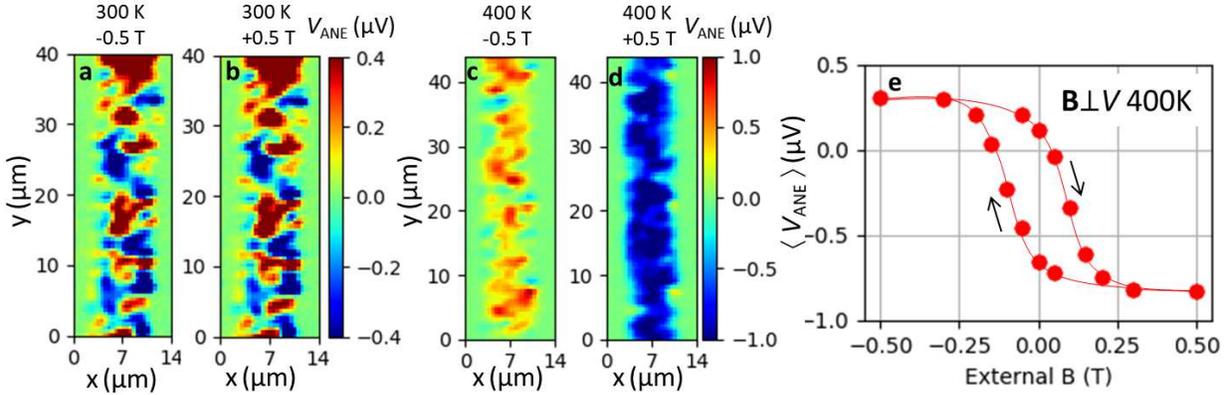}
\caption{Magnetic field dependence of the $V_\text{ANE}$. The magnetic domain structure measured at 300~K does not change upon the application of -0.5~T \textbf{a} and +0.5~T \textbf{b} magnetic field. However, at 400~K, the domain structure can be qualitatively altered by -0.5~T \textbf{c} and +0.5~T \textbf{d}. Panel \textbf{e} shows the average $\left< V_\mathrm{ANE} \right>$ within the scanned region as a function of the magnetic field at 400~K. All scans were taken with a laser power of 10~mW. The red line is a guide for the eye. }

\label{f2}
\end{figure}

We now demonstrate that the $V_\text{ANE}$ signal is indeed of magnetic origin, and that it can be reversed by an external magnetic field. It is important to note that the magnetic field required for the reversal of spins in Mn$_{\text{3}}$Sn thin films is higher compared to bulk Mn$_{\text{3}}$Sn crystals \cite{You2019,Higo2018a}. Therefore, in our thin film samples, the spin structure cannot be reversed by 0.5~T (the experimentally available field) at temperatures below 300~K, as can be seen in Fig.~2a,b. At 375~K a small variation of the measured $V_\text{ANE}$ signal depending on the polarity of the magnetic field  can be seen (SI Fig.~5) corresponding to a reorientation of domains with the weakest coercivity. In contrast, increasing the sample temperature to 400~K and then applying a magnetic field allows us to completely alter the domain pattern, as shown in Fig.~2c,d. At 400~K, the majority of domains is following the external magnetic field. A magnetic field of \SI{-0.5}{T} yields a positive STGM map (red color) across the entire region scanned with the laser spot (Fig.~2c), while for +\SI{0.5}{T}, the STGM map turns  negative (blue color) (Fig.~2d). This shows that at \SI{400}{K}, which is close to the N\'eel temperature $T_\text{N}=\SI{420}{K}$, a magnetic field of \SI{0.5}{T} suffices to align the $\textbf{g}$ vector along the field direction in the entire sample. At the same time the net magnetic moment detected by SQUID magnetometry (see SI, Fig.~2) remains unchanged between 300~K and 400~K, such that  ferromagnet-like phases appearing at higher temperatures can be excluded. To further study the impact of the magnetic field on the magnetic domain pattern, we recorded STGM maps for several different magnetic field values in a ``field sweep''. As detailed in the SI, a complex reversal behavior with multiple domains is observed (for individual maps at each magnetic field see SI Fig.~4). In Fig. 2e, we plot the voltage $\left< V_\mathrm{ANE} \right>$ averaged over the whole scanned area as a function of the field strength, and find a \textit{global} magnetic hysteresis curve with a clear saturation, coercivity and remanence. Since we observe a sign reversal in $\left< V_\mathrm{ANE} \right>$ as a function of the magnetic field, the signal is odd under spin reversal and therefore the main contribution has the Nernst symmetry. The magneto-thermo-voltage clearly cannot be explained by an ordinary Nernst effect, which is linear in magnetic field and does not show hysteresis. Instead, the observed $V_\text{ANE}$ must be connected to the magnetic order parameter of the antiferromagnet. Moreover, when applying the magnetic field parallel to the voltage detection direction, $\left< V_\mathrm{ANE} \right>$ shows no remanence or saturation at \SI{400}{K} (SI Fig.~3). This supports the notion that the  component of \textbf{g} perpendicular to the voltage detection determines $V_\text{ANE}$.

An additional confirmation that the spatial contrast of STGM maps is governed by the antiferromagnetic order in Mn$_{\text{3}}$Sn is evident from the evolution of the STGM signal with temperature.  Figure 3 shows $V_\text{ANE}$ scans taken at different $T$ in the range of \SI{17}{K}  to  \SI{430}{K}. The sample was first cooled to \SI{17}{K} in zero magnetic field and then warmed up step by step to higher temperatures. No magnetic field was applied during this experiment, such that also data at 400~K exhibit a lower voltage $\left< V_\mathrm{ANE} \right>$ compared to the polarized state (Fig.~2). We find a subtle but robust variation with $T$. The net amplitude $\left< V_\mathrm{ANE} \right>$ 
plotted in Fig.~3e peaks in the vicinity of \SI{150}{K},  in excellent agreement with the temperature dependence of the anomalous Nernst response reported in bulk Mn$_{\text{3}}$Sn crystals \cite{Ikhlas2017} which is reproduced in Fig.~3e (blue line). Above $T_\text{N}=\SI{420}{K}$, the thermo-voltage signal vanishes. Indeed, the magnetic field dependent experiments performed at \SI{430}{K}, depicted in the inset of Fig. 3e, show that  $\left< V_\mathrm{ANE} \right>$ is zero, with no signs of remanence or saturation. Note that we did not detect any evidence for a spin glass phase in our samples below 50~K \cite{Nakatsuji2015}. Here again, the different magnetic anisotropy of thin films might affect the presence of the spin glass phase \cite{Ritchey1993,Bisson2008}. 

\begin{figure}[h]
\hspace*{0cm}\epsfig{width=1\columnwidth,angle=0,file=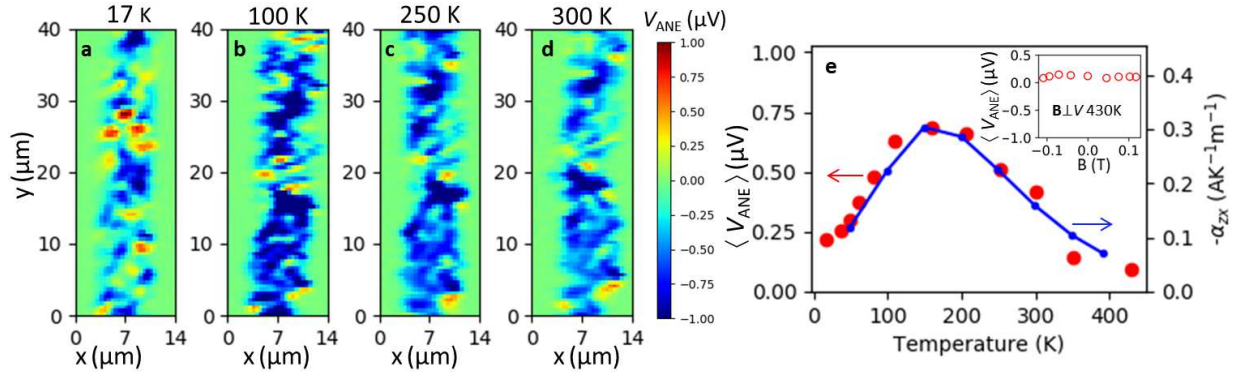}
\caption{Temperature dependency of the measured signal. $V_\text{ANE}$ changes with the sample temperature, as evident from the STGM maps at 17~K \textbf{a}, 100~K \textbf{b}, 250~K \textbf{c}, and 300~K \textbf{d}. All scans were taken with a laser power of 10~mW. \textbf{e} The average $\left< V_\mathrm{ANE} \right>$(T) (red circles) in our thin film exhibits the same temperature dependence as the anomalous Nernst effect reported in the bulk (blue line, data reproduced from \cite{Ikhlas2017}). Inset: Averaged voltage $\left< V_\mathrm{ANE} \right>$ as a function of external magnetic field (equivalent to the experiment in Fig.~2e) at 430~K (above the N\'eel temperature of Mn$_{\text{3}}$Sn). }
\label{f3}
\end{figure}

The STGM does not allow for a direct evaluation of the anomalous Nernst coefficient. The reason is that, unlike in the case of a conventional in-plane thermal gradient, the magnitude of $\nabla T$ cannot be directly measured  since no thermometry ``below the film'' is possible. The laser induced thermal gradient can, however, be estimated by comparing to a film with a known magneto-thermal coefficient deposited on a similar substrate.  
For the present study, we use a thin film of the Weyl semimetal Co$_2$MnGa \cite{Reichlova2018}. We obtain a thermal gradient  of $\nabla T$~$\approx$~2~K$/\mu$m for a laser power of 10~mW, as detailed in the SI, resulting in an anomalous Nernst coefficient of 1.5-2~$\mu$V/K for the Mn$_{\text{3}}$Sn thin film at room temperature. This is higher than the value reported in bulk Mn$_{\text{3}}$Sn crystals 0.6~$\mu$V/K \cite{Ikhlas2017}, which could be due to the higher degree of domain polarization or the more localized detection in our thin film sample. Since the estimation of the thermal gradient contains significant error bars, a more systematic study of the Nernst effect magnitude in thin films is an important future task. 

The capability to intentionally write magnetic domains is exceptionally important not only from an application perspective \cite{Parkin2008}, but also to quantify spin transfer torque \cite{DuttaGupta2015}, domain wall motion or giant magnetoresistance \cite{Gregg1996} in antiferromagnets. In Fig. 4, we show that magnetic domains can be intentionally written into the Mn$_{\text{3}}$Sn film  at room temperature, via a combination of high-power (\SI{50}{mW}) laser illumination and external magnetic fields. Hereby, it is of key importance that the external magnetic field does \textit{not} alter the magnetic texture in Mn$_{\text{3}}$Sn at room temperature in the absence of the laser heating (see Fig.~2a,b).

Fig. 4 shows a sequence of writing and erasing of domains at 300~K using the following procedure: First, the full area (region of interest) is scanned with a \SI{50}{mW} laser spot in +\SI{0.5}{T} external field applied along $x$ direction. Subsequently, the same area is investigated using STGM with \SI{10}{mW} laser power (Fig.~4a), showing a homogeneous thermo-voltage contrast and, thus, homogeneous \textbf{g} vector orientation. In the next step, only the area enclosed by the dashed line is scanned with \SI{50}{mW}, now with applied \SI{-0.5}{T}, followed by imaging the entire area with STGM with \SI{10}{mW} laser power (Fig.~4b). Repeated STGM maps with either \SI{+0.5}{T}, \SI{0}{T} or \SI{-0.5}{T} have no impact on the written pattern (SI Fig.~8), even a magnetic field of 6~T applied along $x$ direction at 300~K could not erase the domain pattern (SI). 

However, increasing the laser power up to \SI{50}{mW} again allows us to write domain patterns at will (Fig.~4c-e). In particular, re-writing the same area with the same magnetic field polarity does not alter the $V_\text{ANE}$ contrast, while writing with opposite field polarity invariably reverses the sign of the thermo-voltage. 

\begin{figure}[h]
\hspace*{0cm}\epsfig{width=1\columnwidth,angle=0,file=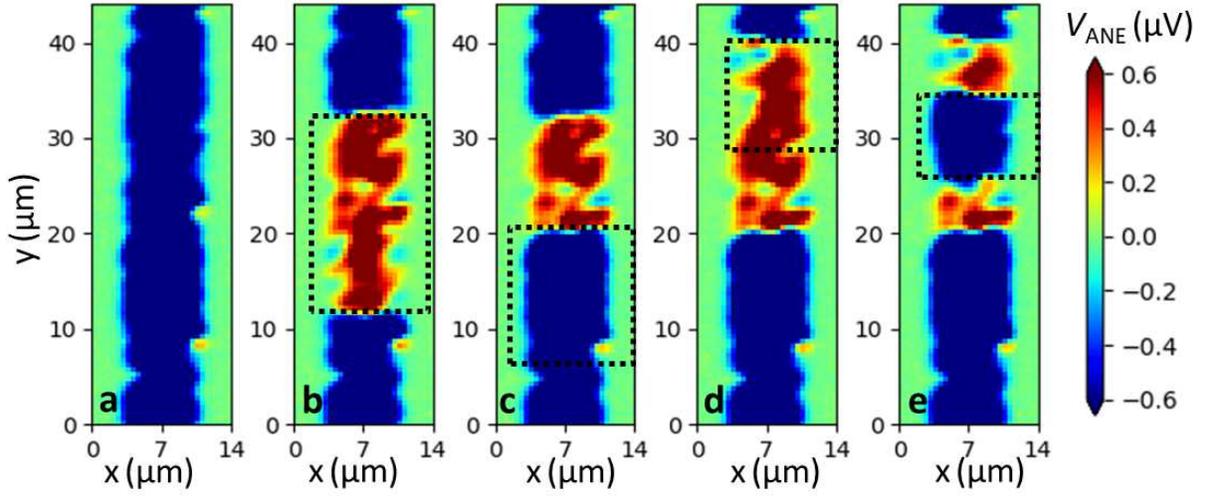}
\caption{Domain writing. Sample polarized to single domain \textbf{a}. In panels \textbf{b-e} the area depicted by a dashed line was written using a laser power of 50mW and the corresponding polarity of 0.5~T magnetic field prior to the STGM scan (+\SI{0.5}{T} yields blue contrast and -\SI{0.5}{T} yields red contrast). Writing was followed by reading with laser power of 10~mW. The whole sequence was performed at room temperature.}
\label{f4}
\end{figure}

Our work demonstrates that the magnetic structure in the non-collinear antiferromagnet Mn$_{\text{3}}$Sn can be spatially mapped out using local laser heating in combination with anomalous Nernst effect measurements. At room temperature and below, the magnetic structure is insensitive to magnetic fields up to 6~T. We further show that domains can be intentionally written into the magnetic structure, via the application of local heat in combination with moderate magnetic fields (\SI{\pm 0.5}{T}). Our experiments, thus, open a pathway to initializing and detecting a domain pattern in non-collinear antiferromagnetic thin films. This represents a first important step towards studying the intriguing physics of non-collinear AFMs with spatial resolution, such as spin transfer torque-induced domain wall motion or spin flop mechanisms in triangular spin systems. Our technique can be  straightforwardly extended to a range of materials, given that they exhibit a finite magneto-thermal response (ANE, anisotropic magneto-thermo power etc.) and, therefore, represents a versatile tool for the investigation of local magnetic properties. 

\pagebreak

Methods: 
\textbf{Growth}
Mn$_{\text{3}}$Sn films were grown by a BesTec UHV magnetron sputtering system on MgO (111) substrates with a 5 nm Ru underlayer. Prior to deposition, the chamber was evacuated to a base pressure of less than 5 $\times$ 10$^{- 9}$~mbar, while the process gas (Ar 5 N) pressure was 3 $\times$ 10$^{- 3}$~mbar. The Ru underlayer was deposited at a rate of 0.45 \AA /s by applying 30 W dc power to a 2~inch target. The Mn$_{\text{3}}$Sn films were grown by cosputtering. The Mn was deposited at a rate of 0.49 \AA /s by applying 42 W dc power and the Sn at a rate of 0.30 \AA /s by applying 11 W dc power to a 3 and 2~inch target, respectively. The growth rates and the film thicknesses were determined by a quartz crystal microbalance and confirmed by using x-ray reflectivity measurements. The substrates were rotated during deposition, to ensure homogeneous growth. The Ru underlayer was  grown at 400$^\circ$~C, the Mn$_{\text{3}}$Sn at RT  and finally the stack was post-annealed at 300$^\circ$~C in-situ for 10 mins. All films were capped with 3~nm of Al to prevent oxidation.
\textbf{Experimental Setup} The thermal gradient is generated by a continuous wave laser operating at a wavelength of $\lambda$=800~nm and focused by an objective lens to a spot size of $\sim$ 1.5~$\mu$m (see Fig.1c). The laser power can be continuously tuned by a combination of a half-wave plate and a polarizer. Scanning of the laser spot across the Hall bar is accomplished by moving the objective lens with a 3D piezo-positioner and the thermo-voltage generated along the $y$ direction is recorded in each position. The laser beam is modulated by a chopper at frequency of 1.7~kHz and the generated thermo-voltage is detected using a lock-in amplifier.

We acknowledge Jakob Lindermeir for technical support, IFW Dresden for providing access to the MST lab,  EU FET Open RIA Grant no. 766566, support from the Grant Agency of the Czech Republic under EXPRO grant no. 19-28375X, and financial support from the DPG through project B05 and C08 of SFB 1143 (project-id 247310070).

%

\end{document}